{
\documentclass[floatfix,prb,twocolumn,nopacs,nofootinbib]{revtex4-2}

\usepackage{amsfonts}
\usepackage{amsmath}
\usepackage[dvipsnames]{xcolor}
\usepackage{nicefrac}
\usepackage{graphicx}
\usepackage{microtype}
\usepackage{physics}
\usepackage{mathrsfs}
\usepackage{amssymb}
\usepackage{bbold}
\usepackage{ dsfont }
\usepackage{float}
\usepackage[bookmarks=true,colorlinks,linkcolor=blue,urlcolor=blue,citecolor=blue]{hyperref}
\usepackage[english]{babel}
\usepackage[applemac]{inputenc}
\usepackage[T1]{fontenc}

\usepackage[export]{adjustbox}
\newcommand{\beq}{\begin{equation}}
\newcommand{\eeq}{\end{equation}}
\newcommand{\beqa}{\begin{eqnarray}}
\newcommand{\eeqa}{\end{eqnarray}}

\usepackage{soul}
\usepackage[normalem]{ulem}

\newcommand{\btext}[1]{\textcolor{blue}{{#1}}}

\graphicspath{{../}}

\begin{document} 

\title{Boomerang effect in classical stochastic models}
\author{Santiago Zamora$^{1}$, Lisan M. M. Dur\~ao$^{2}$, Flavio Noronha$^{1,3}$ and Tommaso Macr\`{i}$^{4,1}$}
\affiliation{
$^{1}$Departamento de F\'{i}sica Te\'{o}rica e Experimental, Universidade Federal do Rio Grande do Norte, Campus Universit\'{a}rio, Lagoa Nova, Natal-RN 59078-970, Brazil\\
$^{2}$ GAVB Consulting,  R. S\~ao Pedro, 440, Canoas - RS, 92020-480, Brazil \\
$^{3}$Institute for Theoretical Physics, Utrecht University, Princetonplein 5, 3584 CC Utrecht, The Netherlands\\
$^{4}$ITAMP, Harvard-Smithsonian Center for Astrophysics, Cambridge, Massachusetts 02138, USA
}

\date{\today}

\begin{abstract}
The phenomenon of Anderson localization, occurring in a disordered medium, significantly influences the dynamics of quantum particles. A fascinating manifestation of this is the "quantum boomerang effect" (QBE), observed when a quantum particle, propelled with a finite initial velocity, reverses its average trajectory, eventually halting at its starting point. This effect has recently been demonstrated in an experiment replicating the quantum kicked-rotor model.
This research delves into the classical analog of QBE. We uncover evidence of a similar effect in classical systems, characterized by the absence of typical diffusion processes. Our investigation encompasses both simplified probabilistic models and more complex phenomenological models that link classical with quantum mechanics. The results indicate that the boomerang effect is not confined to the quantum realm and may also be present in diverse systems exhibiting subdiffusive behavior.
\end{abstract}

\maketitle

\section{Introduction}
Anderson's realization that the interference of the wave function of a particle can stop classical diffusion~\cite{Anderson1958Absence} has led to several interesting results. On the experimental side,  Anderson localization (AL) has been demonstrated in several physical systems such as light~\cite{Chabanov2000, Schwartz2007}, ultrasound waves~\cite{Hu2008}, and atomic matter~\cite{Julien2008, Manai2015, Billy2008, Jendrzejewski2012, Semeghini2015}. Of particular interest to this work is the Quantum boomerang effect (QBE), which was first described in the context of the Anderson model~\cite{Prat2019Quantum}, followed by other models displaying AL~\cite{Tessieri2021Quantum}. Later, it was shown that QBE may appear in non-Hermitian systems~\cite{nonHermitian}, systems without time-reversal symmetry, and sufficient conditions for the existence of QBE were presented~\cite{Janarek2022,PhysRevB.106.L060301,janarek2023berezinskii}.
The QBE has been recently observed experimentally in an implementation of the quantum kicked-rotor model with ultracold bosons~\cite{Sajjad2022}. 
Also, the QBE is destroyed in the presence of mean-field interactions~\cite{Janarek2020Quantum,PhysRevB.106.L060301} and many-body interactions~\cite{JanarekManyBQBE} and it was shown that in a two-component Fermi gas the total wave packet presents a QBE while the individual spin components do not \cite{QBEInFermiones}. 
Whereas the QBE has been deeply investigated in quantum systems,
the goal of this work is to contribute to the understanding of this phenomenon by considering classical particles in a variety of different physical models.

Classical particles may be diffusive in models whose quantum counterparts are Anderson localized. An example of such a situation is a random potential in one dimension (1D). 
This case was investigated for a Gaussian distribution for the disordered potential and, while the quantum model presents AL and QBE, the classical counterpart does not display a boomerang-like effect~\cite{Prat2019Quantum} due to diffusion.

However, the difference between the diffusion in a quantum system and the diffusion in its classical analog is not fundamental. There exist cases where quantum and classical counterparts have similar diffusion behaviors.  A well-known example is the similarity between random walks and Anderson Localization. If one considers a random walk in 1D and 2D, the returning probability to the origin is found to be one.  However, for higher dimensions, the returning probability is less than one. This can be interpreted as the system being more diffusive so the random walker cannot always come back. Similarly, in 1D and 2D, Anderson localization appears for any intensity of disorder in the Anderson model. Nevertheless, for weak disorders in 3D, localization might not be present. The system continues to be diffusive~\cite{Delande_AL_review}. Again, this can be interpreted as the system being more diffusive due to the physical available space. 

This relation between Anderson Localization and random walks, suggests the study of boomerang trajectories in the context of Brownian motion (BM).  The quantum dynamics of a Brownian particle is entirely determined by the dynamical structure factor of the heat bath. In particular, a bath of non-interacting harmonic oscillators with a power-law spectral density has been shown to produce localized states to a positive fractional exponent smaller than one~\cite{leggett1987dynamics}. A bath of two-level systems, also with a power-law spectral density, has been reported to generate insulating behavior and a localization effect in the dynamical sense~\cite{ferrer2006optical,ferrer2007dynamical}. Optical properties of such systems can be studied in linear response using a generalized Langevin equation (GLE)~\cite{lisy2019generalized} and appropriate sum rules. The insulating state can be fully characterized using the Drude weight~\cite{scalapino1993insulator,kohn1964theory}. 
This phenomenological approach, together with the system-plus-reservoir model, brings a formal framework to study boomerang trajectories in both the classical and quantum regimes.

The paper is structured as follows. In Sec.~\ref{sec:ProbModels}, we consider probabilistic models, showing evidence that a classical analog of the boomerang effect (CBE) appears in regimes without diffusion whereas it disappears if the system turns diffusive. 
We obtain approximate analytical expressions for the average dynamics of a particle in a random potential and we finally investigate numerically different situations for this system, finding results that agree with the analytical description. 
In Sec.~\ref{sec:CLTheory} we briefly introduce the relevant formalism of Brownian motion to our study. We review the quantum and classical limits of a Brownian particle immersed in a bath of non-interacting harmonic oscillators in the so-called sub-Ohmic regime. 
The  GLE associated with the problem predicts boomerang-like trajectories in both the quantum and classical limit when the system's variance increases slower than diffusion, which includes the non-diffusive and sub-diffusive regimes.   Interestingly,  in the Ohmic regime which presents normal diffusion, these trajectories do not appear. We present a numerical result for a different memory function, showing that the result aforementioned is not due to a particular description of the bath. Our conclusions are presented in Sec.~\ref{sec:conclusions} where we conclude that boomerang-like trajectories can be expected to appear in either  quantum or classical sub-diffusive systems.

\section{Boomerang Effect in Classical Probabilistic Models}\label{sec:ProbModels}

In this section, we explore a variety of probabilistic models for classical particles in one dimension (1D). In all cases, we consider a Hamiltonian of the form
\begin{equation}
    H = \frac{p^2}{2m} + V(x),\label{Hamiltoniann}
\end{equation}
where $p$ represents the particle's momentum and $V(x)$ denotes a potential characterized by certain randomness. To explore boomerang-like trajectories within classical models, we begin with an analytical examination of a particle's motion through a medium containing thinly dispersed, fixed scatterers, each exhibiting random intensities. This analysis is then juxtaposed with insights from a numerical model, where scatterers are uniformly distributed in space and their magnitudes follow a normal distribution. Furthermore, we simulate numerically the movement of a particle through a random continuous potential in two distinct scenarios: one with a Gaussian-distributed potential and another with a uniformly distributed potential.

In each model, we identify situations where the Classical Boomerang Effect (CBE) appears and scenarios where it does not manifest. Our investigation is divided into two main cases. Initially, we demonstrate that a constant potential at the origin, represented as $V(x=0) = V_0$ in each realization, leads to the emergence of the CBE. In contrast, for Gaussian potentials, we find that choosing $V_0$ from a Gaussian distribution for each realization results in the disappearance of the CBE.

\subsection{Analytical Description}\label{analytical}

We begin by considering a simple model of fixed scatterers that allows us to obtain analytical expressions for the position of a mobile particle. A classical particle of mass $m$ is launched from $x=0$ at time $t=0$ with initial velocity $v_0$ and experiences a constant potential $V_0$ everywhere, except at discrete positions $x_i$ where fixed scatterers are randomly placed. At these positions, the potential varies, and in certain cases, it is sufficiently high to cause the particle to reflect. After this reflection, the particle retains the magnitude of its velocity. We assume that the scatterers are uniformly distributed over all positions.
We define a probability density $f(t')$ indicating the likelihood that the particle will reflect for the first time at time $t=t'$ and at position $x'=v_0t'$.
For a given disorder realization $r$ of the positions and intensities of the scatterers, the particle encounters the first scatterer at $x>0$ after a time $t=t_1$, and another scatterer at $x<0$ after a time $t=t_2$. 
The trajectory is given by 
$x^{(r)}(t;t_1,t_2)=v_0t-2v_0\sum_{n=0}^{+\infty}(-1)^n[t-(n+1)t_1-nt_2] \theta[t -(n+1)t_1-nt_2]$, where the superscript $(r)$ labels the realization.
The averaged trajectory across all disorder realizations with initial potential $V_0$ is
\beqa
 \widetilde{x}(t)=\int_0^{+\infty}\int_0^{+\infty} x^{(r)}(t;t_1,t_2) f(t_1)f(t_2) dt_1 dt_2,\label{eqdensity}
\eeqa 
where $x^{(r)}(t;t_1,t_2)$ is given by the expression above for a single disorder realization.

Given that the scatterers are uniformly distributed over all space, one finds $f(t)=(1/\tau)\,\textrm{e}^{-(t/\tau)}$, where $\tau=\int_0^{+\infty} t f(t)dt$ is the mean free time and the mean free path is $\ell=v_0\tau$ (see Appendix~\ref{app.prob} for details).
After some algebra, one gets
\beqa
\widetilde{x}/\ell&=&2-t_\tau-2\textrm{e}^{-t_\tau}+2\sum_{n=1}^\infty(-1)^n[1+2n\nonumber\\
& &+\,n^2\textrm{e}^{-t_\tau/ n}-(1+n)^2\textrm{e}^{-t_\tau/(n+1)}-t_\tau],\label{xtilde1}
\eeqa 
where $t_\tau=t/\tau$ is a dimensionless time.

\begin{figure}[bt]
\includegraphics[width=0.87\columnwidth,left]{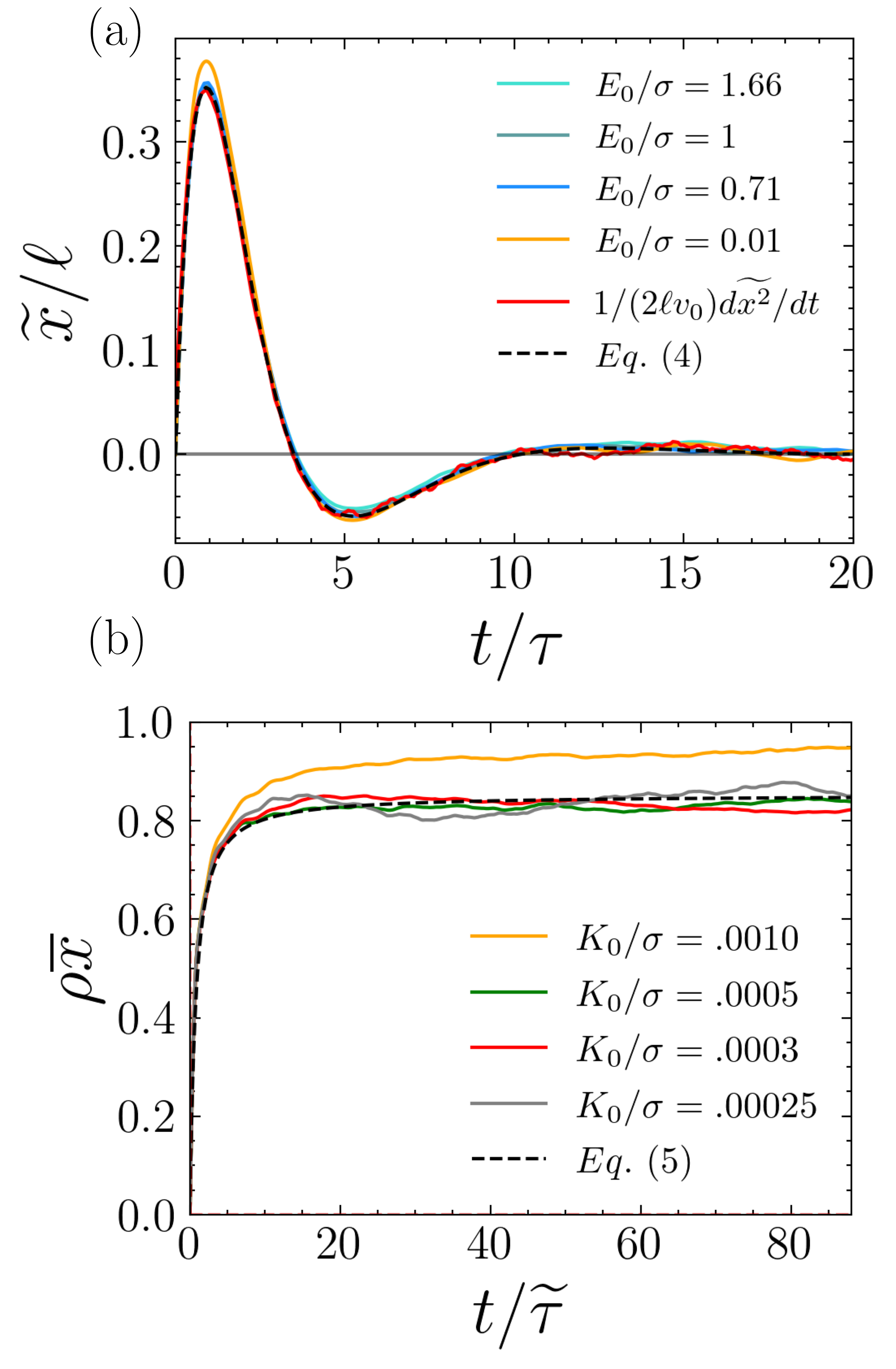}
\caption{\textbf{The Emergence of the Classical Boomerang Effect.} \textbf{(a)}~CBE in the case of a uniform potential $V_0$ with random scatterers. $\tau$ and $\ell$ are the mean free time and path, respectively. The dashed black curve shows the first 60 terms of Eq.~(\ref{bmrV0}) or alternatively the first 900 terms of Eq.~(\ref{xtilde1}). The yellow and the three bluish curves show trajectories obtained from numerical simulations for different values of the potential's variance $\sigma$ and initial energy $E_0$ for the model of scatterers discussed in Sec.~\ref{ec:NumResul_ProbModels}.  The models yield the same average trajectory independently of $\sigma$ and $E_0$ provided ${E_0}/{\sigma}$ is not very small (see main text). The red curve, obtained with $E_0/\sigma=1$, corresponds to the left side of the dynamical relation~(\ref{GeneralDynRel}) in dimensionless units. \textbf{(b)}~Disappearance of the boomerang effect when including randomness in the initial potential $V_0$. $\rho$ is the density of discretization, and $\widetilde{\tau} = \rho v_0$ with $v_0$ being the initial velocity of the particle. The dashed black curve is the plot of  Eq.~(\ref{xGaussian}). The colored plots correspond to the scatterers model with a normally distributed $V_0$ as discussed in Sec.~\ref{ec:NumResul_ProbModels}. Each color corresponds to a different value of $K_0/\sigma$ where $K_0$ is the initial kinetic energy and $\sigma$ is the variance of the potential. We show that for $K_0/\sigma\approx0$ the curves overlap with the analytical prediction. 
} \label{figV0_1}
\end{figure}

We establish the convergence of the series upon assessing the large-$n$ behavior of the terms of Eq.~(\ref{xtilde1}). Similarly, we expand the exponentials for short times and perform the summation over $n$ to obtain:
\beq
\frac{\widetilde{x}(t_\tau)}{\ell}
=4\sum_{l\geq 1}^\infty \frac{(-1)^lt_\tau^l}{l!}(-1+2^{3-l})\zeta(l-2),\label{bmrV0}
\eeq 
where $\zeta(n)$ is the Riemann zeta function. 
There is a divergence in $\zeta(l-2)$ at $l=3$, but the function $(-1+2^{3-l})\zeta(l-2)$ is continuous at $l=3$ and equals to $-\textrm{Log}(2)$. 
The trajectory $\widetilde{x}(t)$ obtained above is shown in the black dashed line of Fig.~\ref{figV0_1}(a) and provides evidence of the existence of a classical boomerang effect as $\lim_{t\to\infty}\widetilde{x}=0$.

\begin{figure}[bt]
\includegraphics[width=0.9\columnwidth,left]{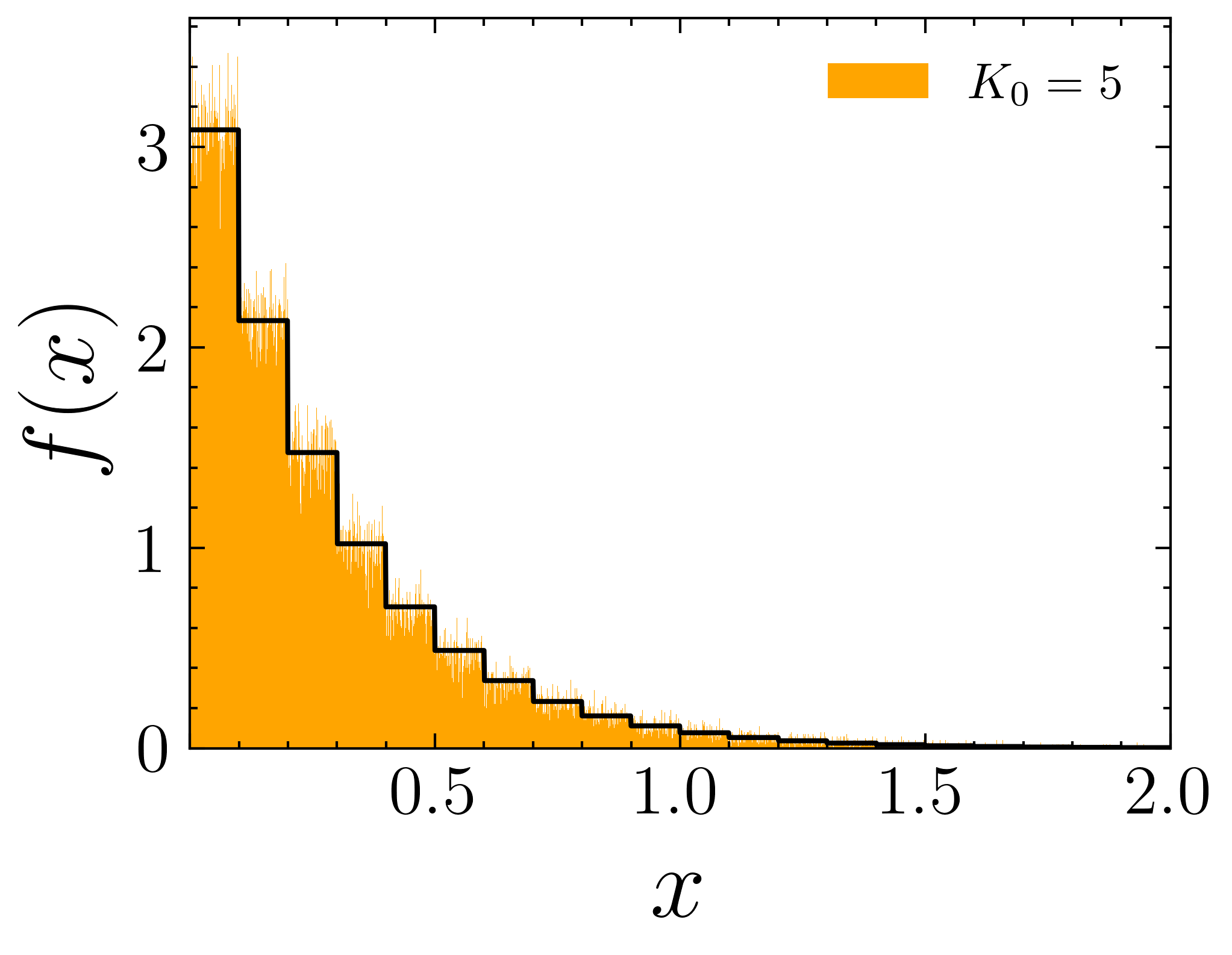}
\caption{\textbf{Scatterer Model Probability Distribution.} This figure illustrates the probability distribution of reflecting at position $x$ in the scatterer model of Sec.~\ref{ec:NumResul_ProbModels}. The histogram, depicted in orange, was compiled from $10^4$ realizations. In each, we recorded the position $x$ where a particle with initial energy $E_0$ first encountered a scatterer with a potential $V(x) > E_0$. Due to the density of scatterers in an interval $\Delta x$ being equal to one in every realization, and their uniform distribution over this interval, plateaus of size $\Delta x$ are observed. In this instance, the variance $\sigma$ of the scatterers' potentials is twice the initial energy of the particle, $\sigma = 2E_0$. The black curve represents the probability distribution $f(x) = \frac{(1-P)}{\Delta x}P^{\lfloor{\frac{x}{\Delta x}}\rfloor}$ of the particle reflecting at $x$ without having reflected before. $P$ is defined in the main text.}
 \label{figV0_2}
\end{figure}

Next, we explore the average trajectory when $V_0$ is sampled from a probability distribution $p(V_0)$ in each realization. This process involves integrating Eq.~(\ref{bmrV0}) with the weight of $p(V_0)$ over all possible values of $V_0$. The integral is solvable for a general $p(V)$ in the scenario where the initial kinetic energy approaches zero, $K_0 \to 0$. Under these conditions, we derive:
\beqa
\rho\overline{x}
&=&\frac{7\zeta(3)}{\pi^2}+\tilde{t}\,^2\bigg\{\textrm{Log}\left[(\Gamma\left(1+\tilde{t}/2)\right)^8/4\pi^2(\Gamma(1+\tilde{t}))^4\right]\nonumber\\
& &+\frac{4}{3}\,\tilde{t}\,\textrm{Log}(2)\bigg\}
-4\zeta'(-2,1+\tilde{t})+32\zeta'(-2,1+\tilde{t}/2)\nonumber\\
& &+8\tilde{t}\left[\zeta'(-1,1+\tilde{t})-4\zeta'(-1,1+\tilde{t}/2)\right],\label{xGaussian}
\eeqa 
where $\Gamma(z)=\int_0^\infty t^{z-1}e^{-t}dt$ is the Euler gamma function, $\zeta(s,a)=\sum_{k=0}^\infty (k+a)^{-s}$ is the generalized Riemann zeta function and $\zeta'(s,a)$ is its partial derivative with respect to $s$.
$\rho^{-1}$ sets a length scale for the system and  $\tilde{t}=t/\widetilde{\tau} = \rho v_0 t$  (see Appendix~\ref{app.prob} for details).
This approximate expression for the average trajectory in the case of a general distribution $p(V)$ corresponds to the dashed black line in Fig.~\ref{figV0_1}(b).
We note that in Ref.~\cite{Prat2019Quantum}, the authors utilized the Boltzmann theory to predict the dynamics of a classical system subjected to a random potential sampled from a Gaussian distribution, resulting in the disorder-averaged position $\overline{x(t)}=\ell'(1-e^{-t/\tau'})$, where $\ell'$ is the scattering mean free path and $\tau'$ is the scattering mean free time.
The trajectory depicted in Fig.~\ref{figV0_1}(b) closely aligns with the result of Ref.~\cite{Prat2019Quantum}, and it is evident that the boomerang effect is absent in both instances, as $\overline{x(t)}$ stabilizes at a finite value when $t\to\infty$.

\subsection{Numerical Results for the Scatterer Model}\label{ec:NumResul_ProbModels}

To assess the generality and accuracy of the previously obtained analytical expressions, we now numerically investigate the average dynamics of a particle moving in several kinds of disordered potential. We start by considering a model of scatterers similar to the one discussed above.
Here we place one scatterer per interval of size $\Delta x$. 
In each disorder realization, the positions of the scatterers are randomly distributed over each of these intervals with a uniform probability distribution. In every position without a scatterer, the potential is $V_0$.
Moreover, a random value for its potential $V(x_j)$ according to a probability distribution $p(V)$ is attributed to each scatterer. 
We define $P=\int_{-\infty}^{V_0+mv_0^2/2}p(V)dV$  as the probability that ${V}(x_j)<{V}_0+mv_0^2/2$, i.e.~the probability that the potential of an unknown scatterer is not high enough to reflect the particle. 
Then the probability distribution of the particle reflecting at some position~$x$, not being reflected in a previous position, has the form $f(x) = [(1-P)/\Delta x]P^{\lfloor{\frac{x}{\Delta x}}\rfloor}$, where $\lfloor{y}\rfloor$ denotes the integer part of $y$. $f(x)$ has the form of an exponential decay with plateaus of size $\Delta x$ (see Fig.~\ref{figV0_2}).
We use a normal distribution $p(V) = e^{-V^2/{2\sigma^2}}/\sqrt{2\pi}\sigma$, for which we find $P = [1+ \textrm{Erf}(E_0/\sqrt 2\sigma)]/2$, where $\textrm{Erf}(x) = (2/\sqrt{\pi})\int_0^xe^{-t^2}dt$ is the error function.
As the ratio $E_0/ \sigma$ decreases, the plateaus become more pronounced, and the distribution $f(x)$ increasingly deviates from the simple exponential decay used for $f(x)$ in the analytical results of Sec.~\ref{analytical}.
Consequently, in the regime where $E_0 \ll \sigma$ \btext{,} the average trajectory of the classical particle is expected to differ from the theoretical prediction of Eq.~(\ref{bmrV0}). 

In Fig.~\ref{figV0_1}(a), we show the average trajectory for four values of $E_0/ \sigma$. We use the dimensionless units $\widetilde{x}/\ell$ and $t/\tau$ defined by the mean free path length $\ell = 1/(\rho\, log(P^{-1}))$ and the mean free time  $\tau = 1/(\rho\, v_0\,log(P^{-1}))$.
The numerical results closely align with Eq.~(\ref{bmrV0}) for most values of $E_0/ \sigma$. Even in the regime $E_0/ \sigma\ll 1$, where the trajectory is expected to differ from the analytical result, we observe reasonable agreement.
Sampling the potential $V_0$ from a Gaussian distribution with variance $\sigma$ in each realization, we replicate the result of Eq.~(\ref{xGaussian}). 
In Fig.~\ref{figV0_1}(b) we show the numerical result (colored solid lines) for four different choices of  $K_0/\sigma$. We see that as 
$K_0/\sigma\rightarrow 0$, the curves overlap with the analytical result (black dashed line).  

We consider the equation of motion of the function $(x^{(r)})^2(t)$ of the $r-$th realization using the  Hamiltonian of Eq.~(\ref{Hamiltoniann})
  \begin{eqnarray}
      \frac{d (x^{(r)})^2}{dt} &=& \{(x^{(r)})^2,H\} \nonumber \\ &=&  \frac{2}{m}x^{(r)}p^{(r)},
 \end{eqnarray}
 where $\{\bullet,\star\}$ denotes the Poisson bracket between $\bullet$ and $\star$.
 where $\{\bullet,\star\}$ denotes the Poisson bracket between $\bullet$ and $\star$.
 Averaging over all realizations one gets, 
\begin{eqnarray}
      \frac{d \overline{x^2}(t)}{dt} 
      =  \frac{2}{m}\overline{px}.
\end{eqnarray}
If the momentum and the position are uncorrelated, then the system satisfies the dynamical relation
\begin{eqnarray}
      \frac{d \overline{x^2}(t)}{dt} 
      =  2 \overline{v} \cdot \overline{x}. \label{GeneralDynRel}
\end{eqnarray}
For systems where $\overline{x^2}(t)\propto t^\alpha$ with $0<\alpha<1$, the dynamics are sub-diffusive. In this case, according to the dynamical equation,  $\overline{x}(t) \propto t^{\alpha-1}\rightarrow 0 $ as $t\rightarrow\infty$, $i.e.$ a CBE appears at long times. 
This indicates a direct relation between the sub-diffusive behavior and the boomerang effect. 
After rescaling the position and time with the characteristic mean free path and mean free time, the dynamical relation simplifies to the more concise dimensionless expression $d \widetilde{x_d^2}/{dt_d} =  2\widetilde{ x}_d $, where $x_d =x/\ell$, $t_d =t/\tau$ and we have used $\overline{v} =l/\tau = v_0$.
In Fig.~\ref{figV0_1}(a) we show the derivative of  $\widetilde{x^2}$ obtained numerically for $E_0/\sigma = 1$ (red curve), which overlaps with the trajectories of the center of mass, demonstrating the validity of the dynamical equation for this model. 

The exploration of the CBE in sub-diffusive systems is deferred to Sec.~(\ref{sec:CLTheory}). In the following subsection, we study the dynamics of a particle moving through a random continuous potential where we explicitly show that classical boomerang trajectories appear whenever the system is not diffusive and disappear when the system displays normal diffusion.

\subsection{Numerical results for a continuous random potential}\label{num.cont}

Here we present the numerical results obtained for a classical particle in 1D moving in a continuous random potential.
In each realization we generated a set of random numbers following a probability distribution $p(V)$ defining the potential at certain positions in the space with a constant spatial density $\rho$. Then we performed an interpolation of these points using the so-called piecewise cubic Hermite interpolating polynomials to generate a smooth (differentiable) potential.   
We then used a standard $4$-th order Runge-Kutta algorithm to solve the dynamical equations.
We studied the case where the potential is sorted from a Gaussian distribution $p(V)$, and the case where it is sorted from a uniform distribution. Once we choose a probability distribution $p(V)$ for the disordered potential and we fix an initial velocity $v_0$ (or kinetic energy $K_0$), we analyze two possible scenarios. In the first scenario, we fix $V_0$ and compute the average trajectory considering only the disorder realizations where $V(x=0)=V_0$. In the second scenario, we extract the average trajectory with all values of $V(x=0)$ allowed by $p(V)$.

\begin{figure}[bt]
\centering   \includegraphics[clip,width=0.9\columnwidth,left]{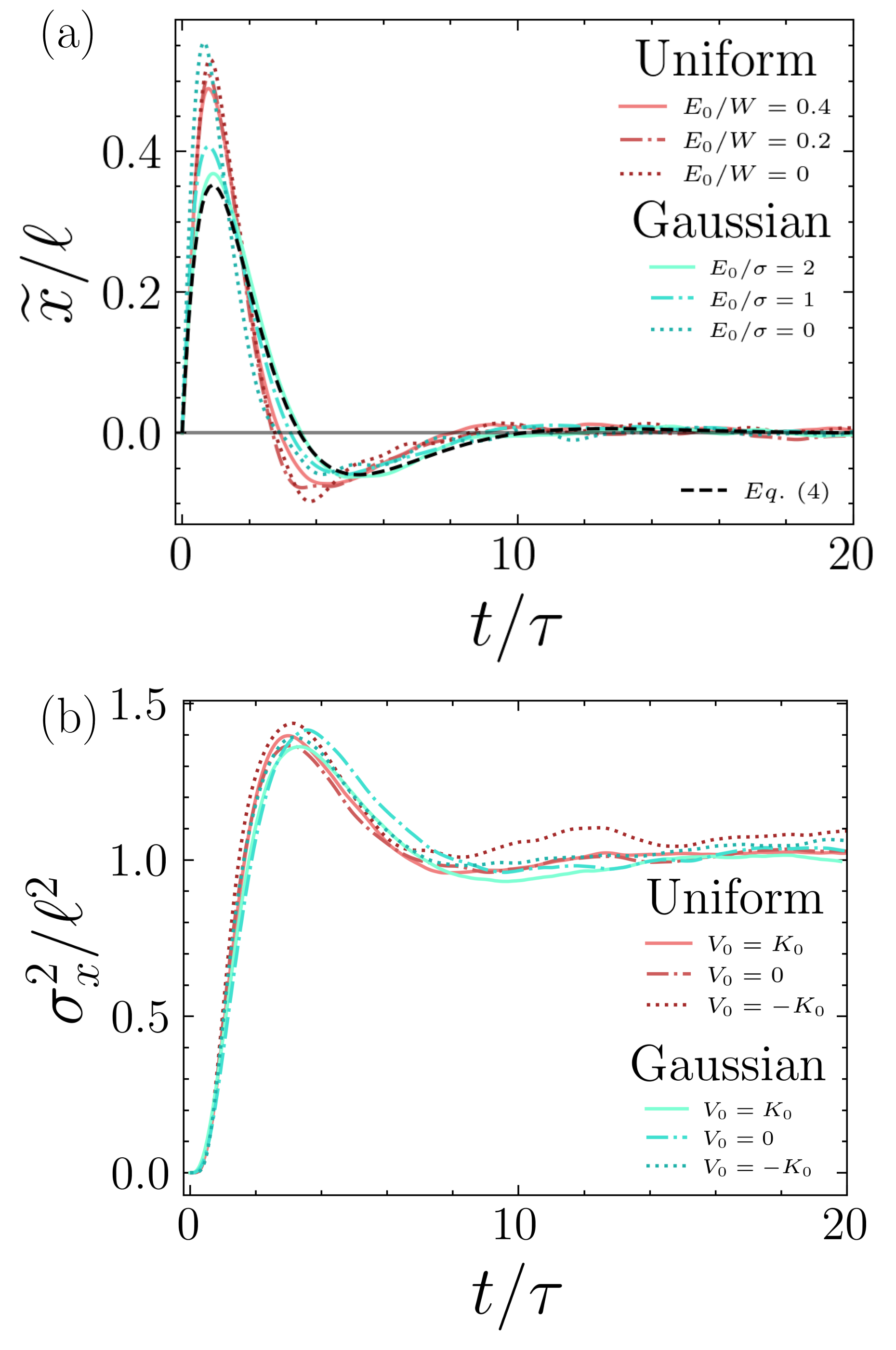} 
\caption{\textbf{Mean trajectories  of a particle with fixed initial energy moving through a continuous random potential.} These plots correspond to the average over $10^6$ realizations. The characteristic units are the mean free path $\ell$ and  the mean free time $\tau$, both calculated numerically.  The initial kinetic energy is $K_0$ in every case. A very important condition is that the potential in the origin is fixed across all realizations, this causes a boomerang-like effect to appear. For the uniform distribution (red curves) the potential is bounded between $\pm W=\pm 5K_0$ and for the Gaussian distribution (cyan curves), the potential is centered in $0$ and has a standard deviation of $\sigma=K_0$.   \textbf{(a)}~Here we show the mean trajectories of an ensemble of particles for the two different distributed random potentials with different fixed potential  $V_0 = V(x_0=0)$ at the origin. The dashed black curve corresponds to Eq.~(\ref{bmrV0}). \textbf{(b)}~The variance of the  mean trajectories shown in (a). We see the system is not diffusive.}
 \label{fig:FixedpotentialV}
\end{figure}

We start investigating the uniformly distributed potential, for which $V(x) \in [-W,W]$, where $W$ is the disorder strength. We fix $V(x=0)=V_0$ and $K_0$, such that $E_0=K_0+V_0$ is fixed. 
If $E_0>W$, the particle will always move forward and will never reflect, the system presents diffusion and it is impossible to obtain a boomerang-like average trajectory.  
However, if $E_0<W$, then in each realization the particle will be trapped in some interval. This allows the CBE to appear, as can be seen in the red curves of Fig.~\ref{fig:FixedpotentialV}(a). The presence of CBE is connected to the absence of diffusion [see Fig.~\ref{fig:FixedpotentialV}(b)].

In contrast to the uniform distribution, with the normal distribution $p(V) = e^{-V^2/{2\sigma^2}}/\sqrt{2\pi}\sigma$ for any fixed $E_0$ there will always be a potential $V_1>E_0$ which will cause the particle to reflect, in such a way each trajectory is always trapped in an interval. As a consequence, for fixed $K_0$ and $V_0$ the average trajectory always displays a CBE and the system is non-diffusive (see the cyan curves in Fig.~\ref{fig:FixedpotentialV}). We plot Eq.~(\ref{bmrV0}), obtained analytically for a scatterer model, in the black dashed curve of Fig.~\ref{fig:FixedpotentialV}(a). It is remarkable that this equation captures the main features of the average dynamics of all the non-diffusive models discussed here.

Moreover, the curves for the Gaussian case seem to differ substantially from the analytical approximation as ${E_0}$ becomes smaller than $2\sigma$. 
In the limit $E_0 \gg 2\sigma $, the potential is too weak over most of the space for the particle to feel it, so it can be approximated as a constant potential. However, since the potential is Gaussian, there is still a non-zero probability of the particle finding $V(x)>E_0$,  so the particle will be reflected at some point. This is similar to a scatterer model, where the particle does not change its velocity between reflections. 
Therefore, it is expected that for $E_0\ge 2\sigma$ the trajectories are well-described by Eq.~(\ref{bmrV0}), while for values $E_0 <2\sigma$ the equivalence between the scatterer and continuous model breaks down.

To conclude our discussion of this first scenario, we note that for the considered potentials, the model satisfies the dynamical equation~(\ref{GeneralDynRel}) by considering the mean velocity of the particle $\overline{v} =\ell/\tau$, given by the mean free path $\ell$ and  the mean free time $\tau$. Notice that in the scatterer model  the mean velocity exactly corresponds to the initial velocity $\overline{v} = v_0$, since the potential is constant between the times $t=0$ and $t=t_1$. Recall that the mean free path  can be calculated numerically as  the average of the distance traveled by the particle between its initial position and the first collision event over all realizations.  The mean free time can be calculated numerically as  the average time  it takes for the particle to travel the distance  between its initial position and the first collision event over all realizations of the system.\\

\begin{figure}[bt]
 \centering  \includegraphics[clip,width=0.9\columnwidth,left]{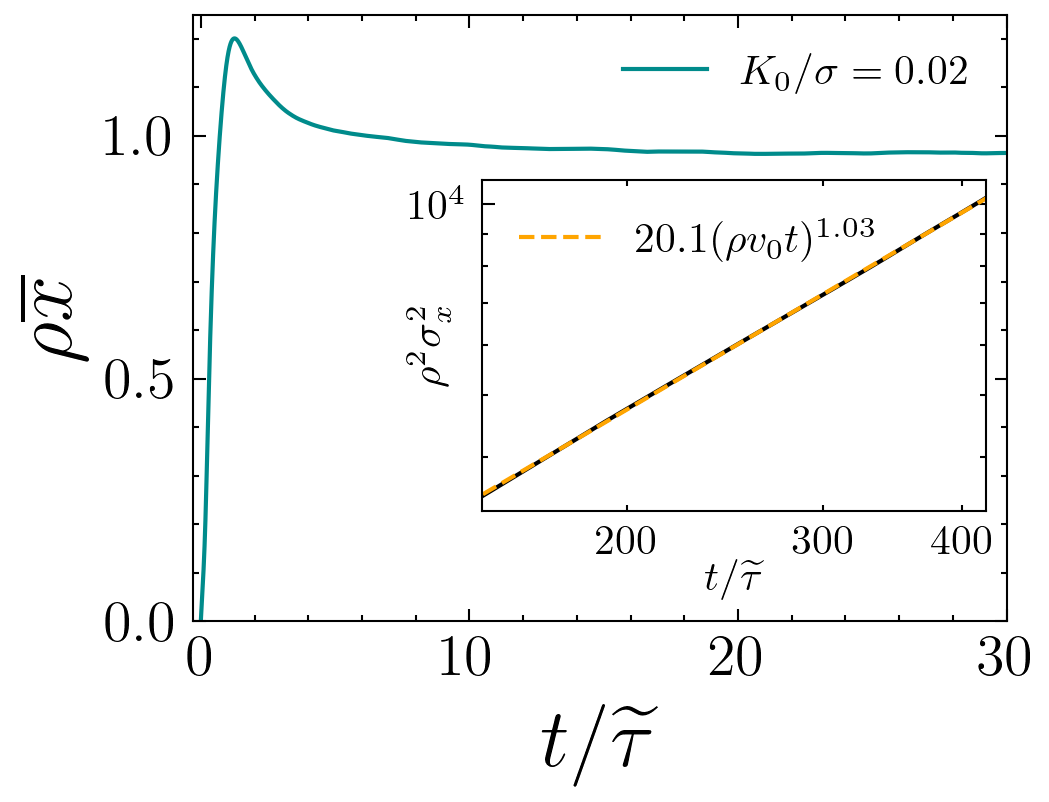} 
\caption{\textbf{Mean trajectory of a particle with random initial energy moving through a continuous random potential.} Average over $10^6$ realizations of a particle moving through a continuous random Gaussian potential in $1$D plotted in dimensionless units where $\rho  $  is the spatial partition density and $\widetilde{\tau} =1/(\rho v_0)$ with $v_0$ the initial velocity of the particle.   In this case the potential in the origin $V_0  $ is random in each realization and the CBE is destroyed.  The variance of the potential distribution is $\sigma = 50K_0 $.  
In the inset, we show the variance of the position of the particle at long times. We observe $\sigma_x^2 \propto t^{1.03}$, indicating normal diffusion.}
 \label{fig:GaussianCurve}
\end{figure}

We now let $E_0$ vary by sorting $V_0$ from a Gaussian distribution for each realization. Interestingly, the boomerang trajectory is lost and we are left with the result shown in Fig.~\ref{fig:GaussianCurve}, which resembles the results in  Fig.~\ref{figV0_1}(b). 
In both cases, as we remove the constraint of fixed $V(x=0)$, the CBE is lost and for large times we find $\rho\overline{x} \sim 1$.
The curves differ for short times, as Fig.~\ref{fig:GaussianCurve} displays a peak. However, the overall behavior is diffusive: 
the inset of Fig.~\ref{fig:GaussianCurve} shows our numerical results for the variance of the system, where one can see that 
$\sigma^2_x(t) \propto t^{1.03}$ for long times. This is evidence of diffusion, which explains the breakdown of the CBE.

We explain the emergence of diffusion in the following. Since the potential has Gaussian distribution, there is always a probability different from zero to have a higher potential than the initial energy of the particle, so it will be reflected at some point with certainty. However, when letting the initial potential to also vary normally in each realization, in some of them the particle will have a very high initial energy. This implies that in these realizations it will take a longer time for the particle to reach a position in space with a potential high enough where it can be reflected. On average, the particle has more available space to move before coming back to the origin, leading to the system's  diffusion and, as a consequence, the boomerang effect disappears.

We emphasize that the CBE is not a trivial phenomenon that emerges when the trajectory is bounded in every disorder realization, as demonstrated by the present case of a disordered potential everywhere. Instead, we note that a boomerang-like average trajectory appears if there is no diffusion at all. The CBE disappears if the system is diffusive.
We recall that through the discussion of the dynamical relation in the previous subsection, we suggested that not only  non-diffusive systems may present a CBE but also systems with sub-diffusion.  This is investigated in the next section using the formalism of general Brownian motion.

\section{Boomerang effect in generalized Brownian motion}\label{sec:CLTheory}
In this section we study the existence of the boomerang effect  in generalized Brownian motion (BM) models. We begin by recalling the general description of both quantum and classical BM, which is given by the generalized Langevin equation (GLE). We review the predicted variance and average position of a particle in the so-called sub-Ohmic regime. 
We then extend the known results to a more general version of the Langevin kernel. 
With this, we show that boomerang trajectories can be found in both quantum and classical models of BM when the system presents sub-diffusion. 
In the following subsections, we first discuss the relevant BM description to our study and then focus on the boomerang trajectories.

\subsection{General description of Brownian Motion}

The system-plus-reservoir approach to quantum BM can be used to describe the center of mass $q(t)$ of a quantum particle moving in a reservoir of quantum harmonic oscillators resulting in the~GLE~\cite{caldeira2014introduction, weiss2021quantum}
\begin{equation}
 M\Ddot{q}(t) + V'(q) +\int_0^{t}\gamma(t-t')\dot{q}(t')dt' = f(t) \label{GLE_with_Vext},
\end{equation}
where $M$ is the mass of the particle and $V'(q)$ is an external force. 
For our purposes, it suffices to take ~$V'(q) = 0$. The memory kernel $\gamma(t)$  describes the influence of the system's past on its present dynamics, while $f(t)$ describes the stochastic force that acts on the system due to the environment. 
The two are related by the fluctuation-dissipation theorem (FDT)
\begin{eqnarray}
    \langle \{f(t),f(t')\}\rangle &=&  \mathcal{F}_c\left[ \gamma(\omega)\hbar\omega \coth\frac{\hbar\omega}{2k_B T}\right], \label{QFDT}
\end{eqnarray}
where $\mathcal{F}_c$ denotes the cosine Fourier transform  and $\gamma(t) = \mathcal{F}_c\left[\gamma(\omega)\right]$ for $t\ge0$ \cite{ghosh2023quantum}. Here $\hbar$ is the reduced Planck constant, $k_B$ is the Boltzmann constant and $T$ is the temperature of the reservoir.

In this context,  it is customary to employ the phenomenological memory function 

\begin{equation}
\gamma(\omega) = A_s\, \omega^{s-1},\label{Phen_Damp_func}
\end{equation}
where the constant $A_s$ has the dimensions $[A_s] = M/T^{2-s}$ and $s\in[0,2]$.  Three regimes appear depending on the value of $s$. For $0\le s< 1$ the regime is called sub-Ohmic while for  $ 1< s \le2 $  it is called super-Ohmic. The Ohmic regime $s=1$ is the frequency independent case which reproduces the usual Brownian motion~\cite{caldeira2014introduction, weiss2021quantum,Uhlenbeck}. The Ohmic regime has normal diffusion and does not present boomerang trajectories \cite{Prat2019Quantum}. 
On the other side, the sub-Ohmic regime displays sub-diffusive behaviour \cite{weiss2021quantum,Resta_2018, scalapino1993insulator, Bertini_FTT, kohn1964theory}.

Although Eq.~(\ref{GLE_with_Vext}) is the equation of motion of the center of mass of a quantum particle moving in a reservoir of quantum harmonic oscillators,  it also describes the trajectory of a classical particle moving in  a bath of classical harmonic oscillators. That is, the average value of the position of the quantum particle follows the trajectory of the classical particle, which is the content of Ehrenfest's theorem \cite{caldeira2014introduction,weiss2021quantum}. As a consequence, the average trajectories that emerge in the quantum case will also appear in the classical regime. In particular, in the next subsection, we show the emergence of boomerang trajectories in both scenarios.  From the quantum description, the classical regime can be obtained by taking the operators in Eq.~(\ref{GLE_with_Vext})  to be classical numbers and considering the limit  of  high temperatures or small frequencies, $\hbar\omega/(k_BT)\rightarrow 0$, so the FDT reads
\begin{eqnarray}
    \langle f(t)f(t')\rangle &=& k_B T \gamma(t-t') . \label{FDT}
\end{eqnarray}

To study the diffusion in the sub-Ohmic regime, we consider the mean square displacement (MSD) of the quantum system $\sigma_Q$ defined as
\begin{equation}
    \sigma_Q^2(t) = \langle \left(q(t)-q(0)\right)^2\rangle.
\end{equation} 
One can show that~\cite{weiss2021quantum}
\begin{equation}
\sigma_Q^2(t) = \frac{\hbar}{\pi}\int_{-\infty}^{\infty}d\omega \chi''(\omega) \coth\left(\frac{\hbar\beta\omega}{2}\right)\left[1-\cos(\omega t)\right],\label{QuantuVar}
\end{equation}
 where $\beta = (k_BT)^{-1}$ and $\chi''(\omega)$  is the imaginary part of the dynamical susceptibility of the Brownian particle corresponding to Eq.~(\ref{GLE_with_Vext}):
\begin{equation}
\chi(\omega) = -\frac{1}{M\omega^2+ i\omega\gamma_F(\omega)}.
\end{equation}
 The sub-index $F$ denotes that $\gamma_F(\omega)$ is the Fourier transform of $\gamma(t)$. 
In general, $\chi(\omega) = \mathcal{F}(\theta(t)\chi(t))$ is the half-Fourier transform  of the response function $\chi(t)$ of the system,  $\theta(t)$ is the Heaviside function.
Considering the limit of  high temperatures and small frequencies, the classical limit of Eq.~(\ref{QuantuVar}) reads
\begin{equation}
\sigma_C^2(t)  = \frac{2k_{B}T}{\pi}\int_{-\infty}^{\infty}\frac{d\omega}{\omega}\chi''(\omega)\left[1-\cos(\omega t)\right]. \label{ClassVar}
\end{equation}
These equations are true for any valid $\gamma_F(\omega)$ and in general the integrals are solvable only numerically. However, for the memory function written in Eq.~(\ref{Phen_Damp_func}) analytical approximations can be made. In the following subsection, we discuss these results and those for the average position of the particle. 

\subsection{Boomerang trajectories in the sub-Ohmic regime.}
In this subsection, we focus on the average trajectory and MSD of the Brownian particle for the memory function given in Eq.~(\ref{Phen_Damp_func}).  In Ref.~\cite{weiss2021quantum} an analytical study shows that $\sigma_Q^2(t) \propto t^{s}$ for long times and $0 \le s<2$. This shows in particular that in the sub-Ohmic regime the system is sub-diffusive. In the classical limit, this behaviour holds and an equivalent result can be found for  $\sigma_C^2(t)$ in Ref.~\cite{wang1992long}.

In Appendix~\ref{AppendixCaputo}, we derive that in this same regime  the average trajectory $\overline{x}(t)$ for both quantum and classical limit is
\begin{equation}
    \overline{x}(t) = v_0 t E_{2-s,2}(-A/M t^{2-s}), \label{AvGPos_ML}
\end{equation}
where $E_{\alpha,\beta}(y)=\sum_{k=0}^{\infty}y^{k}/\Gamma(\alpha k+\beta)$ is the two parameter Mittag-Leffler (ML) function, $\alpha,\beta\in\mathbb{C}, Re(\alpha)> 0,Re(\beta)> 0$ \cite{MittagLeffler_Khan}. $\Gamma(z)=\int_0^\infty t^{z-1}e^{-t}dt$ is the Euler gamma function. In the derivation of Eq.~(\ref{AvGPos_ML}) it was used that in the sub-ohmic regime the memory kernel of the GLE has the form 
\begin{equation}
    \gamma(t) = \frac{2}{\pi}A_s \Gamma(s)\, \cos(\frac{\pi s}{2})\,t^{-s}. \label{GammasForTheSOmhicCLassicalCase}
\end{equation}
\begin{figure}[bt]

 \centering  \includegraphics[clip,width=0.95\columnwidth,left]{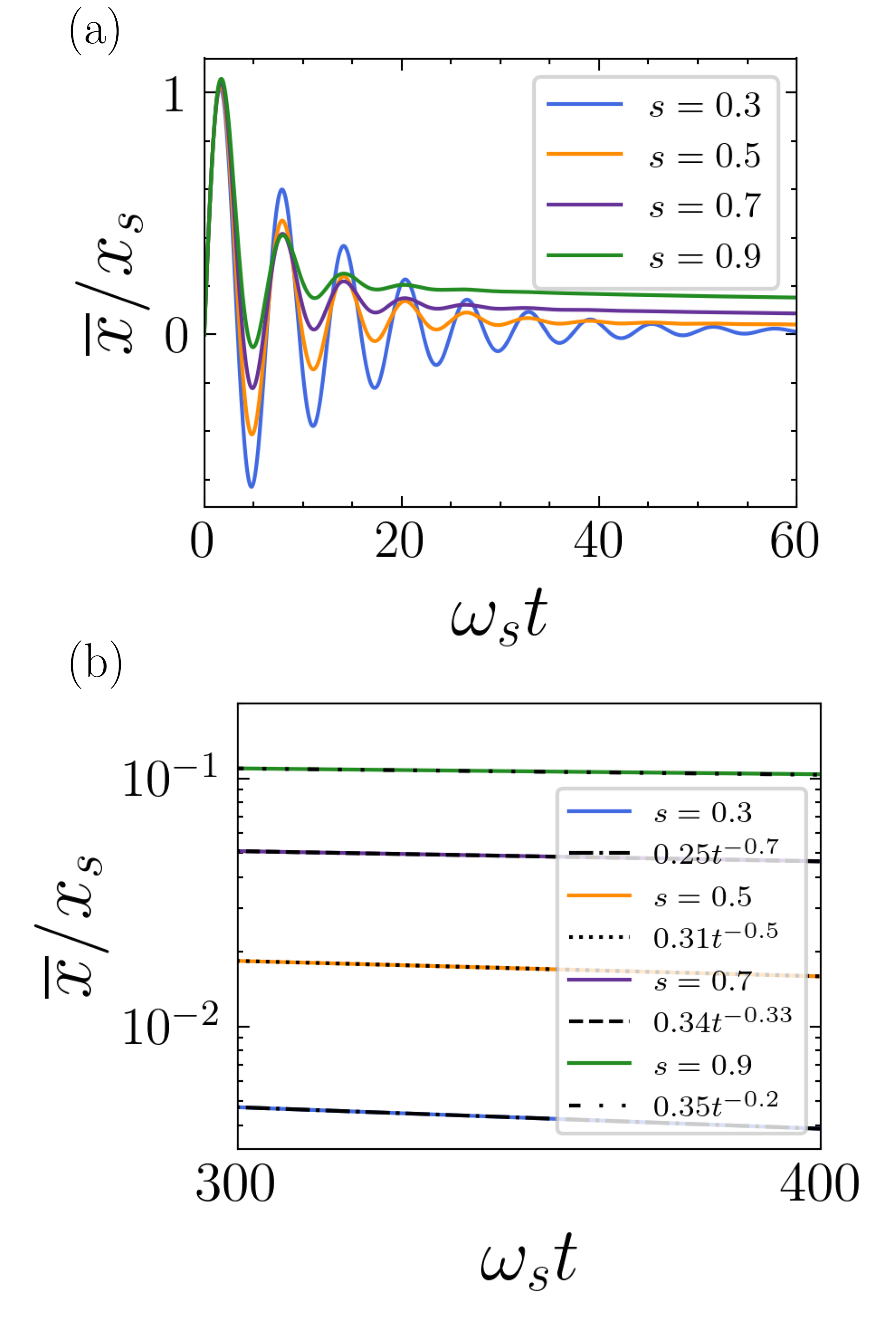}  
\caption{\textbf{Average position of a particle moving in a bath of classical harmonic oscillators.}  Average trajectory of a particle moving in a bath of harmonic oscillators described by the kernel~(\ref{regular_Kernel}) in the sub-Ohmic regime. 
The characteristic frequency and length are given by $\omega_s = {(2A_s\Gamma(s)cos(\pi s/2)/(\pi M))}^{1/(2-s)}$ and $x_s=1/w_s\sqrt{kT/M}$.   \textbf{(a)}~Mean trajectories of the ensemble for different values of $s$. \textbf{(b)}~The curves show a power-law behaviour, approaching $\overline{x}= 0$ at long times.}
\label{fig:NonDivergentKernel}
\end{figure}
Using the asymptotic expansion  at large $y$ of the ML function to first order~\cite{Erdelyi_TranscendentalF},

\begin{equation}
    E_{\alpha,\beta}(y)= \frac{-y^{-1}}{\Gamma(\beta-\alpha)},
\end{equation}
allows us to find the following expression for the long-time average position
 \begin{equation}
 \overline {x}(t)=   \frac{v_0 M}{A\Gamma(s)}t^{s-1}, \label{andamentoTempolongo}
\end{equation}
which displays that $\overline {x}(t)\rightarrow 0$ as $t\rightarrow\infty$. 
This proves the existence of boomerang-like trajectories when the system is sub-diffusive in both the quantum and classical case. 
Importantly, we recall that when $s=1$, the memory function~(\ref{Phen_Damp_func}) is frequency independent and the Ohmic case is recovered. In the Ohmic regime the system does not present boomerang trajectories and displays normal diffusion \cite{caldeira2014introduction,weiss2021quantum}.

\begin{figure}[bt]
 \centering  \includegraphics[clip,width=0.9\columnwidth,left]{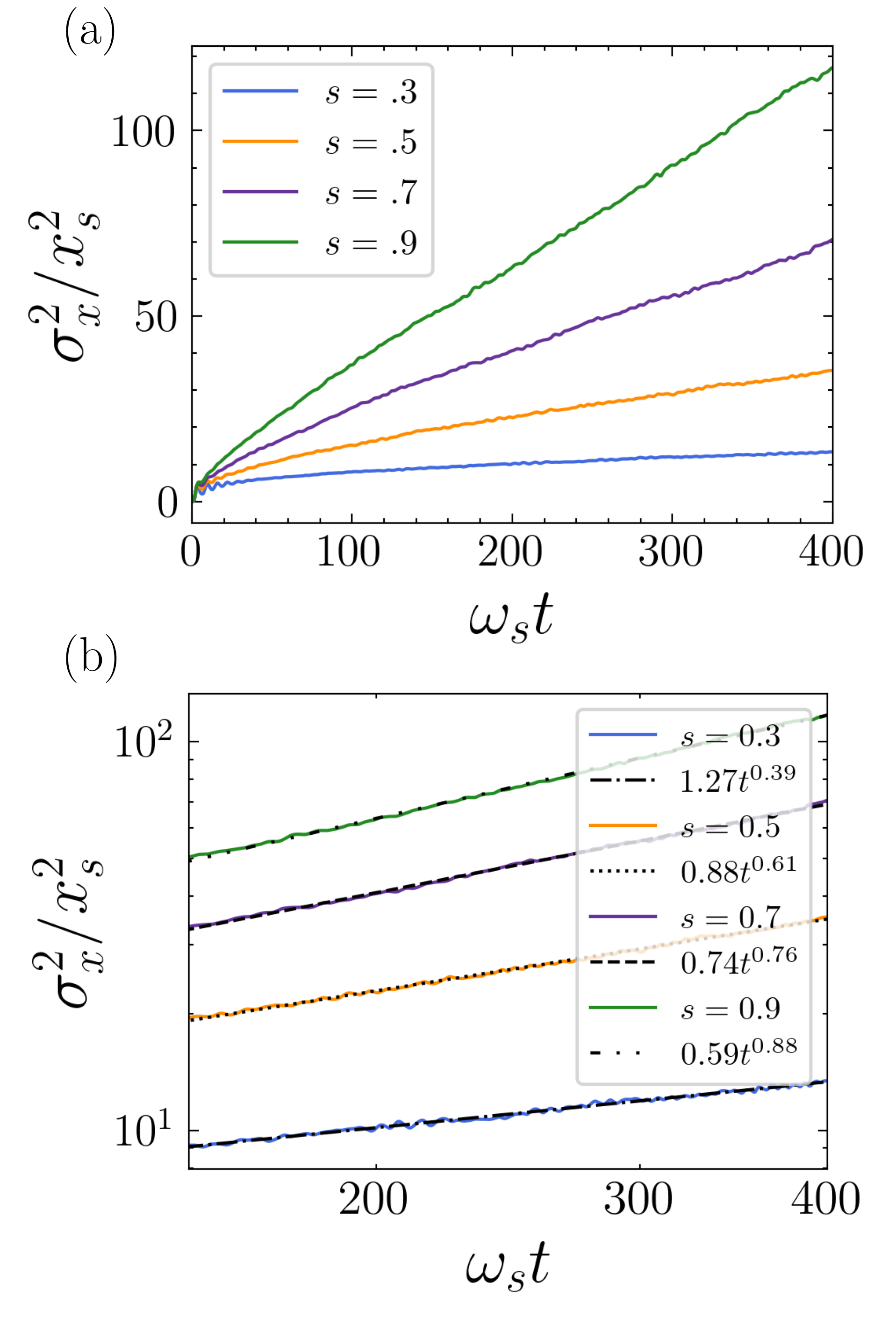}  
\caption{\textbf{Variance of the position of a particle moving in a bath of classical harmonic oscillators.} 
Variance of the trajectories of a particle moving in a bath of harmonic oscillators described by the Kernel~(\ref{regular_Kernel}) in the sub-Ohmic regime. The characteristic frequency and length are given by $\omega_s = {(2A_s\Gamma(s)cos(\pi s/2)/(\pi M))}^{1/(2-s)}$ and $x_s=1/w_s\sqrt{kT/M}$. \textbf{(a)} Variance of the position for different values of $s$ \textbf{(b)} At long times, the variance follows a power-law with an exponent less than 1.  Then the system presents boomerang-like trajectories while being sub-diffusive (see Fig.~\ref{fig:NonDivergentKernel}).}
\label{fig:NonDivergentKernel_var}
\end{figure}

The result for the sub-diffusive case is not only limited to the memory function given in Eq.~($\ref{GammasForTheSOmhicCLassicalCase}$). 
As an example, we consider the classical case with a  regularized kernel at short times defined as

\begin{equation}
    \gamma_r(t) = \frac{2}{\pi}A_s \Gamma(s)\, \cos(\frac{\pi s}{2})\,(1+t^2)^{-s/2}. \label{regular_Kernel}
\end{equation}
With this kernel, a numerical solution of the corresponding GLE is possible. To numerically generate a time series satisfying the corresponding FDT (Eq.~(\ref{FDT})), we used the Fourier filtering method described  in~\cite{PrakasetAl, Makse_LongRangeCorrelations,Spitzner_CorrelatedDisorder}. The results for the mean trajectory are shown in  Fig.~\ref{fig:NonDivergentKernel}. 
We plot the curves in units of the characteristic frequency $\omega_s = (2A_s\Gamma(s)cos(\pi s/2)/(\pi M))^{1/(2-s)}$ and the characteristic position $x_s = v_c/\omega_s$ with $v_c = \sqrt{k_BT/M}$.   
For values below $s=0.5$, the average position has the same power-law behavior proportional to $ t^{s-1}$ shown by the long-time solutions with the divergent kernel. Increasing the exponent $s$, this result seems to hold no more as shown for $s=0.7$ and  $s=0.9$. 
The results of the variance are presented in Fig.~\ref{fig:NonDivergentKernel_var}. 
Note that for considered value $s$, the mean position vanishes for large times while the variance shows the system to be sub-diffusive.
In summary, these results show that boomerang-like trajectories appear in sub-diffusive systems, regardless of whether the system is quantum or classical. 

Before we conclude, we would like to mention Ref.~\cite{ferrer2007dynamical}. In this article, the authors study the effect of a two-level systems reservoir on single particle dynamics. They use the Feynman-Vernon formalism to obtain the equation of motion of the center of mass. 
By considering a similar memory function to the one presented in (\ref{Phen_Damp_func}), they show that the particle center of mass localizes in the sub-Ohmic regime as a function of the temperature. For small finite temperatures and $s$ close to zero, the center of mass of the particle oscillates in a way that resembles that of a particle confined by an effective potential. As the temperature or $s$ is increased, the localization effect gets weaker. The consequence of this dynamical localization is shown to be boomerang-like trajectories of the center of mass.

\section{Conclusions}\label{sec:conclusions}
In this work, we investigated the emergence of a boomerang effect in a broad class of classical systems. We studied classical probabilistic models, consisting of a particle moving in $1$D through random scatterers normally distributed or through a random continuous potential. In the latter model, we considered two cases: We sampled the potential from a uniform and a Gaussian distribution. 
We obtained an analytical expression for the mean trajectory of a particle moving through a homogeneous potential everywhere except at discrete random positions where scatterers of random potential are placed. The analytical expression agrees qualitatively with all the aforementioned models and it is quantitatively a very good description for the scatterers and the continuous Gaussian potential numerical models in the regime of high enough initial energies. 
We also showed that the models satisfy a dynamical equation, which relates the first moment of the position with the time derivative of the second moment of the position through the mean velocity of the particle.
In the continuous random potential models, the mean velocity is determined by the mean free path and mean free time of the system. In the scatterer models, the mean velocity simplifies into the initial velocity of the particle. 
As before, in the limit of high energies the scatterer and the continuous Gaussian potential model satisfy remarkably well the corresponding dynamical equation.

We found the existence of boomerang-like trajectories in all the situations. However, it was found that the appearance of this type of trajectory in the normally distributed continuous random potential case is dependent on the initial energy of the particle. If the energy of the particle is random at every realization, a normal diffusive motion is obtained and no CBE is observed. If both the kinetic and the potential energy at the initial position are fixed in every realization, the variance of the particle's position saturates to a constant value indicating the absence of diffusion and a classical boomerang effect emerges. In fact, in every case where a boomerang-like average trajectory appeared, the absence of diffusion was present.  This seemed to suggest that these boomerang-like trajectories had a relation with the system's diffusion. We conjectured that if the system's variance grows slower than normal diffusion, boomerang trajectories can emerge. To evaluate this hypothesis, we studied sub-Ohmic systems in the context of generalized Brownian motion.

In the sub-Ohmic scenario, we showed that boomerang-like trajectories appear in both the quantum and classical cases while the system is sub-diffusive. In contrast, in the Ohmic regime, the system presents normal diffusion and no boomerang-like trajectories appear.
To demonstrate the generality of the result, we numerically solved the GLE with a modified memory kernel for the classical case. As expected, the particle describes boomerang trajectories on average, and the system is confirmed to be sub-diffusive.

In summary, we showed that the boomerang effect extends beyond the quantum regime of Anderson-localized systems.
Instead, it seems to depend on the diffusion presented by the system: for sub-diffusive systems, including systems with the absence of diffusion, boomerang trajectories may appear. This is consistent with the QBE where Anderson Localization serves as a mechanism to stop diffusion.  

\section*{Acknowledgements}
We extend our gratitude to Patrizia Vignolo for her insightful discussions. This work is dedicated to the memory of Dominique Delande, whose engaging conversations on the boomerang effect greatly inspired and motivated us. 
We thank the High-Performance Computing Center (NPAD) at UFRN for providing computational resources. T. M. acknowledges the hospitality of ITAMP-Harvard where part of this work was done.
This work was supported by the Serrapilheira Institute (grant number Serra-1812-27802). 
This study was financed in part by CNPq and the Coordena\c{c}\~ao de Aperfei\c{c}oamento de Pessoal de N\'ivel Superior - Brasil (CAPES) - Finance Code 001.

\begin{appendix}

\section{Deriving expressions for the probabilistic models}\label{app.prob}

\noindent 
In this Appendix we derive some expressions used in Sec.~\ref{analytical}. The trajectory of a classical particle subject to reflections in scatterers is
\begin{eqnarray}
    x(t;t_1,t_2)&=&vt-2v\sum_{n=0}^{+\infty}(-1)^n[t-(n+1)t_1-nt_2]\nonumber \\
    &\times& \theta[t -(n+1)t_1-nt_2],
\end{eqnarray}
where $t_1$ and $t_2$ are the times described in the main text, $v$ is the initial velocity and $\theta[t]$ is the heaviside function.
Inserting last equation in  {Eq}.~(\ref{eqdensity}) we find, after some algebra, that 
\begin{eqnarray}
    \tilde{x}(t)&=&vt-2vt\int_0^t\bigg(1-\frac{t'}{t}\bigg)f(t')dt'-2vt\sum_{n=1}^\infty (-1)^n \nonumber \\
    &\times&\int_0^t \bigg[\bigg(1-\frac{t'}{t}\bigg){F}\bigg(\frac{t-t'}{n}\bigg)f\bigg(\frac{t'}{n+1}\bigg)\frac{1}{n+1} \nonumber \\
    &-& \frac{t'}{t}{F}\bigg(\frac{t-t'}{n+1}\bigg)f\bigg(\frac{t'}{n}\bigg)\frac{1}{n}\bigg]dt',\label{x_integrals}
\end{eqnarray}
where ${F}(t)=\int_0^t f(t')dt'$ is the probability that the particle will reflect before the time $t$. In the following, we find an expression for $f(t)$ and $F(t)$ to compute $\tilde{x}(t)$. 

We define $P=\int_{-\infty}^{V_0+mv^2/2}p(V)dV$  as the probability that ${V}(x_j)<{V}_0+mv^2/2$, where $p(V)$ is a general probability density for the potential $V(x)$. Therefore the probability that the particle will reflect at the $j$-th position, not being reflected before, is $f(t_j)={P}^{j-1}(1-{P})$. {Then} ${F}(t_j)=1-{P}^j$, where $j=\rho vt_j$ and $\rho={N/L}$ is the density of the discretization, i.e.~the number of positions ${N}$ divided by the length of the space ${L}$. Now we assume that in the continuum limit we have ${F}(t)=1-{P}^{\rho vt}$ and hence $f(t)=\rho v \textrm{Log}({P}^{-1}){P}^{\rho vt}$.  Hence the mean free time is $\tau=\int_0^{+\infty} t f(t)dt=1/\rho v \textrm{Log}({P}^{-1})$ and the mean free path is $\ell=v\tau=1/\rho \textrm{Log}({P}^{-1})$.  
We get $f(t)=(1/\tau)\,\textrm{e}^{-(t/\tau)}$. Notice that $\tau$ depends on $P$ and hence depends on the initial potential $V_0$. For this reason, the trajectory $\tilde{x}(t)$ also depends on $V_0$. With these considerations, one can arrive to Eq.~(\ref{xtilde1}) by performing the integrals in Eq.~(\ref{x_integrals}).

To find an expression for the average trajectory considering all possible values for the initial potential $V_0$, one may compute 
\beqa
\overline{x(t)}=\int_{-\infty}^{+\infty} \tilde{x}(t;V_0) p(V_0) dV_0, \label{xmeanvv}
\eeqa
where again $p(V_0)$ is the probability distribution of the potential $V(x=0)=V_0$ at the initial position $x=0$.
 
Using Eq.~(\ref{bmrV0}) in Eq.~(\ref{xmeanvv}) and writing the dimensionless time as $\tilde{t}= \rho vt$ we find
\beqa
\rho\overline{x}
&=&\sum_{l=1}^\infty \frac{(-1)^l\tilde{t}\,^l}{l!}4(-1+2^{3-l})\zeta(l-2)\mathcal{I}_l(K),\nonumber\\
\mathcal{I}_l(K)& =&\int_{-\infty}^{+\infty}  [\textrm{Log}(P^{-1})]^{l-1}p(V_0)dV_0.\label{xttilde}
\eeqa 
To derive an analytical expression for $\mathcal{I}_l(K)$ we define $u=P(V_0+K)$ and assume $p(V_0)\approx p(V_0+K)$, which is exact in the limit $K\to 0$. Then,
\beqa 
\mathcal{I}_l(K)
& \approx&\int_{0}^{1}  [\textrm{Log}(u^{-1})]^{l-1}du = (l-1)!\,\,.\label{IlK}
\eeqa 

Notice that this expression is valid for any distribution $p(V)$, including the Gaussian $p(V)=\textrm{Exp}(-V^2/\sigma^2)/(\sigma\sqrt{\pi})$, the Lorentzian, etc. 

Replacing Eq.~(\ref{IlK}) in~(\ref{xttilde}) leads to Eq.~(\ref{xGaussian}) in the main text.

\section{Analytical solution of the GLE for the mean displacement }\label{AppendixCaputo}

The equation to solve is 
\begin{equation}
   M \ddot{\overline{x}}(t)+ \int_0^{t}\gamma(t-t') \dot{\overline{x}}(t') dt' = 0, \label{GLE_Mean_value}
\end{equation}
with $\gamma(t-t') =2/\pi A_s  \cos \left(\frac{\pi  s}{2}\right) \Gamma (s)(t-t')^{-s} $ for  $0<s<1$.
Applying the Laplace transform we obtain
\begin{equation}
    z\overline{v}_L(z)-\overline{v}(0) + \gamma_L(z)\overline{v}_L(z)/M = 0,
\end{equation}
where the notation $f_L(z)$  represents the Laplace transform of $f(t)$ and $\overline{v}=\dot{\overline{x}}$. We also used the convolution theorem on the integral term. This gives  the solution in the $z$-space for the velocity 
\begin{equation}
    \overline{v}_L(z) = \frac{\overline{v}(0)}{z+\gamma_L(z)/M}.
\end{equation}
Defining  $B = 2/\pi A_s  \cos \left(\frac{\pi  s}{2}\right) \Gamma (s)$,  the Laplace transform of the kernel $\gamma(t)$ reads
\begin{equation}
    \gamma_L(z) =\int_0^\infty e^{-zt}Bt^{-s}dt = B\Gamma(1-s)z^{ s-1} = Az^{s-1} .
\end{equation}
Replacing it in the expression for the velocity gets us the result
\begin{equation}
    \overline{v}_L(z) = \frac{\overline{v}(0)}{z+Az^{s-1}/M}. \label{VelocityLaplaceSpace}
\end{equation}
From \cite{Mittag-Leffler_functions}, the Laplace transform of the two-parameter Mittag-Leffler function is given by 
\begin{equation}
    L\Bigl(t^{\beta-1}E_{\alpha,\beta}(\lambda t^{\alpha})\Bigr)=\frac{1}{z^{\beta}-\lambda z^{\beta-\alpha}},
\end{equation}
so by setting $\lambda = -A/M$, $\beta = 1$ and $\alpha = 2-s$ one can find 
\begin{equation}
   \overline{v}_L(z) = v_0L\Bigl({E_{2-s,1}(-A/M t^{2-s})}\Bigr).
\end{equation}
Then the solution for the velocity is
\begin{equation}
    \overline{v}(t) = v_0 E_{2-s,1}(-A/M t^{2-s}).
\end{equation}
To obtain the mean position one can use the series definition of the Mittag-Leffler function 
\begin{equation}
    E_{\alpha,\beta}(y)=\sum_{k=0}^{\infty}\frac{y^{k}}{\Gamma(\alpha k+\beta)}.
\end{equation}
Integrating term by term in the velocity, taking a $t$ out of the sum  and using that $((2-s)k +1)\Gamma((2-s)k +1) = \Gamma((2-s)k +2) $, the average position is
\begin{equation}
    \overline{x}(t) = v_0 t E_{2-s,2}(-A/M t^{2-s}).
\end{equation}

\end{appendix}
\newpage
\bibliography{Reference}

\end{document}